\documentclass[letterpaper,twoside]{article}
\usepackage{latexsym,times}
\usepackage[intlimits]{amsmath}
\usepackage{amsfonts,amssymb}
\DeclareSymbolFontAlphabet{\mathbb}{AMSb}
\usepackage[mathscr]{eucal}
\usepackage{changebar} 
\usepackage[textstyle,cdot,squaren,thickspace,thickqspace]{SIunits}
\usepackage[colon,authoryear]{natbib}
\usepackage{fancyvrb}
\usepackage{url}
\usepackage{xspace}
\usepackage[usenames,dvips]{color}
\usepackage[dvips]{graphicx}
\DeclareGraphicsExtensions{.eps}   
\graphicspath{
  {./}
}
\usepackage[letterpaper,top=1in,bottom=1in,hmargin={1in,1in}]{geometry}
\numberwithin{equation}{section}
\begin{document}
\title{Cognitive Maps of Complex Systems Show Hierarchical Structure and Scale-Free Properties}
\author{Uygar \"{O}zesmi$^{1}$ and Can Ozan Tan$^{2,3,*}$}
\date{}
\maketitle

\noindent
$^1$Department of Environmental Engineering, Erciyes University, 38039 Kayseri, Turkey, uygar@ozesmi.org

\noindent
$^2$Department of Biology, Canakkale Onsekiz Mart University, 17020 Canakkale, Turkey.

\noindent
$^3$Department of Cognitive and Neural Systems, Boston University, Boston, MA 02215, USA

\noindent
$^*$Corresponding author, e-mail: tanc@cns.bu.edu

\begin{abstract}
Many networks in natural and human-made systems exhibit scale-free properties and are small worlds. Now we show that people's understanding of complex systems in their cognitive maps also follow a scale-free topology ($P_k = k^{-\lambda}, \, \lambda \in [1.24,3.03]$; $r^2 \geq 0.95$). People focus on a few attributes, as indicated by a fat tail in the probability distribution of total degree. These few attributes are related with many other variables in the system. Many more attributes have very few connections. The scale-free properties in the cognitive maps of people arise despite the fact that their average distances are not different (Wilcoxon sign-rank test, $W=78$, $p=0.75$) than random networks of the same size and connection density. The scale-free property manifests itself in the higher hierarchical structure compared to random networks (Wilcoxon sign-rank test, $W=12$, $p=0.03$). People use relatively short explanations to describe systems. These findings may help us to better understand people's perceptions, especially when it comes to decision-making, conflict resolution, politics and management.
\end{abstract}


\noindent
\Large I\normalsize{n} nature many networks from protein interaction \cite{1} to metabolism \cite{2} show scale-free properties.  In human systems scale-free properties have been observed in the world-wide web \cite{3}, the internet \cite{4}, linguistics \cite{5}, sexual contacts \cite{6}, movie actor collaboration \cite{7}, and scientific collaboration \cite{8}. Here we show that people's perceptions of complex systems, such as ecosystems, also obey power laws.

We examined people's perceptions of complex systems with the technique of cognitive mapping. Cognitive maps are networks that have weighted and directed edges (causal connections of varying strengths) between nodes (variables). Axelrod \cite{9} first used signed digraphs to represent causal relationships among variables as defined and described by people and he called these representations cognitive maps. By using weighted connections instead of binary ones, Kosko \cite{10} defined fuzzy cognitive maps. Cognitive maps have been used to examine decision-making \cite{11}, people's perceptions of complex social systems \cite{12}, and for modelling in various fields including operations management \cite{13}, virtual reality \cite{14} and environmental management \cite{15}. Eden et al. \cite{16} have extensively used cognitive mapping to examine decision-making and problem-solving in businesses. Recently, fuzzy cognitive maps created with expert knowledge have been used in data mining of the world wide web \cite{17}. 

While studying twelve entirely different complex systems through cognitive maps, we have observed an arising pattern of similarity. We asked the question whether these systems exhibit quantifiable and generalizable common patterns with respect to their statistical mechanics.

\section*{Methods}

\noindent
\textbf{Obtaining cognitive maps.} Cognitive mapping is an established methodology applied in many different fields of study. These studies are compared and evaluated, and the system of interviewing and analysis are described and reviewed in detail by \"{O}zesmi and \"{O}zesmi \cite{19}. In this study, we used in-depth interviews with people about various complex systems. In these studies, first the process of drawing a fuzzy cognitive map was shown with a completely unrelated map. Then the stakeholders were asked to draw their own cognitive map in response to open-ended questions such as "What variables come to mind if I mention -insert a system-, how do these variables affect each other?". The stakeholders listed all the variables that came to their minds. They signified the relationships between these variables by drawing lines between them and using arrows to indicate the directions of the relationships. They also gave them signs of positive or negative, and strengths of a lot (1), some (0.5), or a little (0.25). After the stakeholders drew their cognitive maps, which are essentially directed weighted graphs, they were coded into adjacency matrices. These adjacency matrices were then augmented and added together \cite{18} to create a social cognitive map of stakeholder groups or of all the stakeholders interviewed for each study system.

\noindent
\textbf{Structure of cognitive maps.} We analyzed the structure of 12 cognitive maps by examining the outdegree, the indegree, and total degree (centrality) of the variables.  Outdegree shows the total strength of the connections exiting from a variable:

\begin{equation*}
\mathrm{od}(v_i) = \sum_{k-1}^N \bar{a}_{ik}
\end{equation*}

\noindent
Indegree shows the total strength of the connections coming into a variable:

\begin{equation*}
\mathrm{id}(v_i) = \sum_{k-1}^N \bar{a}_{ki}
\end{equation*}

\noindent
Total degree (centrality) is the sum of indegree and outdegree of a variable:

\begin{equation*}
c_i = \mathrm{td}(v_i) = \mathrm{od}(v_i) + \mathrm{id}(v_i) 
\end{equation*}

Total degree shows the cumulative strength of connections entering into and exiting from a variable.  It indicates how important a variable is in the map.

We examined the average distance between variables assuming that the connections are undirected.  

We also looked at other structural indices to determine if maps from different study systems were similar.  We calculated the ratio of receiver to transmitter variables ($R/T$).  Receiver variables have a positive indegree, $\mathrm{id}(v_i)$, and zero outdegree, $\mathrm{od}(v_i)$. Transmitter variables have a positive outdegree, $\mathrm{od}(v_i)$, and zero indegree, $\mathrm{id}(v_i)$. By definition of complexity, maps will have larger ratios of receiver to transmitter variables as their complexity increases, because they define more utility outcomes and less controlling forcing functions.

We examined the density (clustering coefficient) of the fuzzy cognitive maps ($D$), calculated as the number of connections divided by the maximum number of connections possible between $N$ variables \cite{20}: 

\begin{equation*}
D = \frac{k}{N^2}
\end{equation*}

Density is an index of connectivity that shows how connected or sparse the maps are. If the density of a map is high then then there are a large number of causal relationships among the variables.

Another structural measure of a cognitive map is the hierarchy index ($h$) \cite{21}:

\begin{equation*}
h = \frac{12}{(N-1)N(N+1)} \sum_i \left[ \mathrm{od}(v_i) - \frac{\sum \mathrm{od}(v_i)}{N} \right]^2
\end{equation*}
 
\noindent
where $N$ is the total number of variables.  When $h$ is equal to zero the system is fully democratic and when $h$ is equal to one then the map is fully hierarchical.

Average path distances ($l$) of the maps were obtained using Dijkstra algorithm \citep{dijkstra}. Random average path distances ($l_{rand}$) were obtained by the same algorithm from a randomly connected network with the same connection probability and the same size. 

Graph indices were also calculated for each individual cognitive map from a random graph of the same size, same connection probability and same distribution of connection weights. The mean values after 300 monte carlo simulations of the indices calculated from these random maps are compared with the indices obtained from actual cognitive maps for statistical significance using Wilcoxon sign-rank test.

In order to determine the linear part of the exponential distribution of total degrees of the maps, we have fitted linear models to log-log distribution of the total degrees, individually for each map, recursively by pruning the last point from the linear model, until the slope of the linear model stabilized. Exponential distribution coefficients ($\lambda$) and $R^2$ values are obtained by fitting a exponential distribution to the linear part of each map obtained by aforementioned linear models.

\section*{Results}

\noindent
\textbf{Structure of cognitive maps.} Our analysis revealed that the structural indices of the social cognitive maps from separate study systems are in the same range (Table \ref{tab1}).  On average, there are $5.23\pm1.64$ SD connections per variable in the maps. The ratio of receiver variables to transmitter variables ($R/T$) had a median of 0.362, with a range of $[0,15]$. This indicates that numbers of transmitter variables were mostly higher than numbers of receiver variables in the maps (Table \ref{tab1}). Higher numbers of transmitter variables show that people perceived the systems as being governed by forcing functions rather than utility (receiver) variables. Densities of the maps are low ($0.048\pm 0.027$ SD), indicating maps with relatively few relationships among variables.  The low hierarchy index (median: $0.035$, range: [0.001,0.010]) of the maps indicates that the maps tend to be more "democratic" than "hierarchical" in their structure.

\begin{figure}
\includegraphics[width=\textwidth]{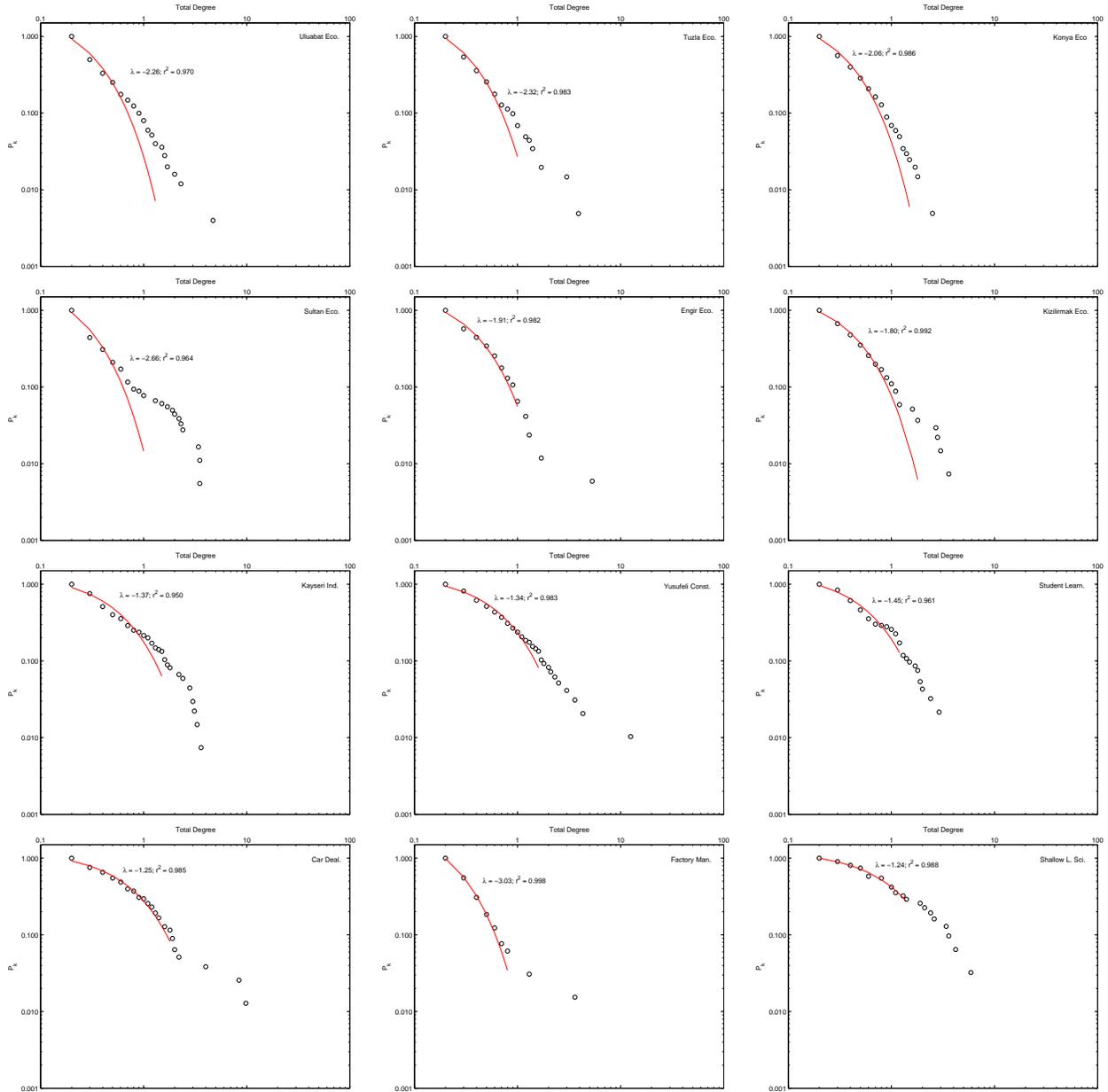}
\caption{Cumulative probability distributions of total degree of social cognitive maps for each complex system considered in this study. Red lines show the exponential fits to log-log distributions of the total degrees ($P_k = k^{-\lambda}$, $\lambda \in [1.24,3.03]$). Note that each social cognitive map shows a fat tail characterizing scale-free networks with power law distribution.}
\label{fig1}
\end{figure}

\begin{figure}
\includegraphics[width=\textwidth]{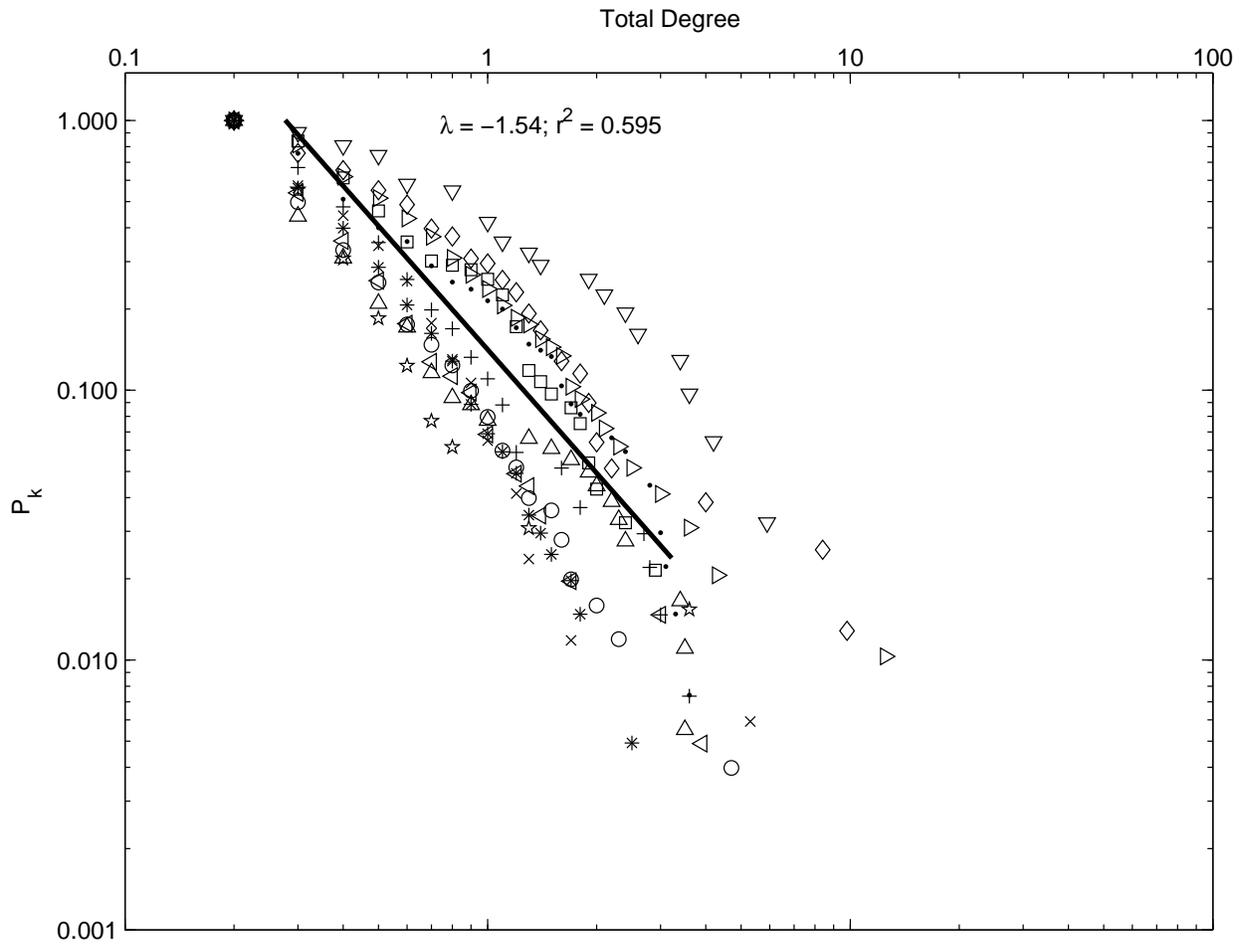}
\caption{Cumulative distributions of pooled total degree from all social cognitive maps ($n=12$), where the exponential fit is $P_k =k^{-\lambda}$, $\lambda=1.54$, $r^2 =0.60$. While each complex system is denoted by a different symbol, a fat tail in the distribution is still evident.}
\label{fig2}
\end{figure}

\noindent
\textbf{Total degree.} When we examined the total degree of the variables in individual stakeholder cognitive maps, they had power, Poisson, bimodal or uniform distributions. However, in social cognitive maps, the degree distribution of the total degree, or centrality, always follows a power law (Figures \ref{fig1} and \ref{fig2}). The total degree exponents, $\gamma_{\mathrm{total}}$, vary from -1.236 to -3.034 for our 12 studies with $R^2$ between 0.950 to 0.998 (Table \ref{tab2}). Because cognitive maps have directed links (i.e. agricultural runoff causes lake eutrophication), we examined the probability densities of the indegree and outdegree. These also followed a power law distribution for the social cognitive maps. The indegree exponents, $\gamma{_\mathrm{in}}$, vary from -1.534 to -6.514 ($R^2$ = 0.965 - 0.999). The outdegree exponents, $\gamma_{\mathrm{out}}$, vary from -1.370 to -14.99 ($R^2$ = 0.974 - 0.998) (Table \ref{tab2}).

\noindent
\textbf{Average distance.} As in most real world networks \cite{22} social cognitive maps are small worlds, with an average map distance ($l$) between nodes varying from 1.94 to 4.92 (Table \ref{tab2}). These distances are similar to the distances (range: $[2.49,3.47]$) in random networks of the same size (Wilcoxon sign-rank test, $W=78$, $p=0.7507$). Although the causal links between variables are directed, the average distances were calculated assuming that the links between nodes were bidirectional. Therefore we would expect the lengths of causal chains of reasoning to be even shorter on average. In other words, people's perceptions of complex systems involve causal chains that are on average less than 5 links long. One could assume that the longer the causal chain the greater level of detail that people use to explain an event.

\section*{Discussion}

Many large networks have the property that the vertex connectivities follow a scale-free power-law distribution because: 1) networks expand continuously by the addition of new vertices, and 2) new vertices attach preferentially to sites that are already well-connected \cite{7}. With social cognitive maps, as more individual maps are added to the social map, the map expands with new nodes and new connections following the same logic. Because people in the same geographic area or social group tend to have similar perceptions about systems or problems, they think of the same variables and causal relationships between those variables. Therefore the more maps that are added together, the stronger the causal connections become between shared variables.

Weighed bidirectional cognitive maps show distinctly different statistical properties compared to random networks of the same average map distance, density, and size. The way people perceive a complex system is significantly  more hierarchical than comparable random networks. People tend to have more forcing functions than utility variables, whereas in random networks these tend to be equal in number. These findings show that the perception of complex systems and the way people describe them is focused on system drivers and ordered causal chains. The causal chains lead to utilities of concern to them. Theoretically these cognitive maps do not fully describe the ``reality'' out there, however models derived using cognitive maps and their simulated results of the ``real'' world  have been found to validate empirical results \cite{tanozesmi2006} and have shown to be useful in decision making (15;16;24-28). 

Social cognitive maps were found to be small worlds with variable connectivities following a scale-free power-law distribution. These results have implications for how people perceive systems and could help solve problems in many different areas such as cognition, perception, decision-making, conflict resolution, politics, and management. For example, by analyzing networks of social interactions together with cognitive maps we may be better able to understand how ideas spread and become accepted by group members. Finally cognitive maps of people's perceptions of complex systems provide another example of networks with scale-free topology, presenting a case where the network connections are directed and weighted.

\small
\noindent
This research was in part funded by the Turkish Scientific and Technical Research Council (TUBITAK-YDABAG), the Turkish Society for the Protection of Nature (DHKD), Turkish State Hydraulic Works, and the University of Minnesota MacArthur Program.

\normalsize


\newpage

\begin{table}[ht]
\begin{center}
\caption{Values for the number of individual maps in the social cognitive map ($n$), number of variables ($N$), number of edges or causal connections ($k$), and graph theory indices, receiver to transmitter ratio ($R/T$), density ($D$), hierarchy ($h$), for social cognitive maps from six different study systems. Values in parentheses indicate the mean value obtained by 300 Monte Carlo simulations from a random network of the same size and equal connection probability.}
\label{tab1}
\vspace{0.1in}
\begin{tabular}{p{1in}lllllll}
System&$n$&$N$&$k$&$k/N$&$R/T^{\S}$&$D$&$h^{\ddag}$\\
\hline
Uluabat Lake \par ecosystem&35&253&1173&4.64&0.116 (1.0053)&0.018&0.011 (0.000016)\\
Tuzla Lake \par ecosystem&44&204&864&4.24&0.391 (1.0133)&0.019&0.024 (0.000016)\\
Konya basin \par ecosystem&152&203&1861&9.17&0.217 (0.9999)&0.045&0.001 (0.000021)\\
Sultan Marshes \par ecosystem&56&181&773&4.27&1.032 (1.0099)&0.023&0.118 (0.000040)\\
Engir Marshes \par ecosystem&43&171&1049&6.14&1.039 (1.0099)&0.036&0.002 (0.000087)\\
Kizilirmak Delta \par ecosystem&31&136&616&4.62&0.333 (1.0090)&0.033&0.026 (0.000089)\\
Kayseri \par industry&30&135&948&7.02&0.640 (1.0021)&0.050&0.080 (0.000054)\\
Yusufeli dam \par construction&14&97&360&3.71&0.083 (1.0128)&0.039&0.049 (0.000029)\\
Student learning&24&95&557&5.86&1.000(1.0110)&0.062&0.017 (0.000094)\\
Car dealerships&26&78&442&5.67&0.000(1.0125)&0.074&0.029 (0.000337)\\
Factory \par management&28&65&256&3.99&15.00 (1.0122)&0.062&0.056 (0.000323)\\
Shallow Lake \par scientists&8&32&113&3.53&0.167 (1.0539)&0.114&0.005 (0.003401)\\
\hline
\end{tabular}
\end{center}
\vspace{0.1in}

\hspace{0.3in}
$^{\S}$ Significantly different from random network (Wilcoxon sign-rank test, $W=17;p=0.0923)$.

\hspace{0.3in}
$^{\ddag}$ Significantly different from random network (Wilcoxon sign-rank test, $W=12;p=0.0343)$.
\end{table}

\newpage
\begin{table}[ht]
\begin{center}
\caption{Values for the number of individual maps in the social cognitive map ($n$), number of variables ($N$), number of edges or causal connections ($k$), and total degree ($\lambda_\mathrm{total}$), indegree ($\lambda_\mathrm{in}$) and outdegree ($\lambda_\mathrm{out}$) exponents with the corresponding $R^2$ values for power law distribution together with the average distances, l, of social cognitive maps from six different study systems. Values in parentheses ($l_\mathrm{rand}$) indicate the mean value obtained by 300 Monte Carlo simulations from a random network of the same size and equal connection probability. ``Ref'' indicates published results of individual cognitive mapping studies.}
\label{tab2}
\begin{tabular}{p{1in}llllllllllll}
System&$n$&$N$&$k$&$\lambda_\mathrm{total}$&$R^2$&$\lambda_\mathrm{in}$&$R^2$&$\lambda_\mathrm{out}$&$R^2$&$l (l_\mathrm{rand})^{\ddag}$&Ref.\\
\hline
Uluabat Lake \par ecosystem&35&253&1173&-2.258&0.970&-4.125&0.979&-5.373&0.991&3.77 (3.41)&\cite{21}\\
Tuzla Lake \par ecosystem&44&204&864&-2.324&0.983&-4.325&0.993&-4.775&0.990&4.92 (3.47)&\cite{24}\\
Konya basin \par ecosystem&152&203&1861&-2.055&0.986&-4.817&0.986&-4.733&0.998&2.60 (2.59)&--\\
Sultan Marshes \par ecosystem&56&181&773&-2.663&0.964&-6.098&0.980&-5.251&0.989&4.69 (3.33)&\cite{23}\\
Engir Marshes \par ecosystem&43&171&1049&-1.913&0.982&-3.857&0.989&-3.139&0.999&2.78 (2.85)&\cite{engir}\\
Kizilirmak Delta \par ecosystem&31&136&616&-1.803&0.992&-3.783&0.961&-2.765&0.994&3.72 (3.20)&\cite{19,18}\\
Kayseri \par industry&30&135&948&-1.367&0.950&-2.125&0.965&-2.044&0.971&3.06 (2.63)&\cite{22}\\
Yusufeli dam \par construction&14&97&360&-1.344&0.983&-2.095&0.984&-1.790&0.986&3.780 (3.17)&\cite{20}\\
Student learning&24&95&557&-1.453&0.961&-2.165&0.990&-1.994&0.986&2.63 (2.65)&--\\
Car dealerships&26&78&442&-1.248&0.985&-2.274&0.982&-1.457&0.974&2.56 (2.60)&\cite{coban:auto}\\
Factory \par management&28&65&256&-3.034&0.998&-6.514&0.999&-14.99&0.997&1.94 (2.89)&\cite{coban:factory}\\
Shallow Lake \par scientists&8&32&113&-1.236&0.988&-1.534&0.977&-1.370&0.991&2.27 (2.59)&\cite{tanozesmi2006}\\
\hline
\end{tabular}
\end{center}

$^{\ddag}$ Not significantly different from random network (Wilcoxon sign-rank test, $W=78;p=0.7507)$.
\end{table}

\clearpage

\clearpage

\end{document}